\documentclass[useAMS,usenatbib]{mn2e}

\usepackage{amssymb}
\usepackage{amsfonts}
\usepackage{mncite}
\usepackage{epsfig}
\usepackage{psfig}

\begin{document}

\title[Evolution of hot Jupiters]{Why are there so few hot Jupiters?}

\author[W.K.M. Rice, Philip J. Armitage \&
 D.F. Hogg]{W.K.M. Rice$^1$\thanks{E-mail: wkmr@roe.ac.uk}, Philip
 J. Armitage$^{2,3}$,
D.F. Hogg$^4$ \\
$^1$ SUPA\thanks{Scottish Universities Physics Alliance},
Institute for Astronomy, University of Edinburgh, Blackford Hill, Edinburgh, EH9 3HJ \\
$^2$ JILA, Campus Box 440, University of Colorado, Boulder CO 80309 \\
$^3$ Department of Astrophysical and Planetary Sciences, University of
 Colorado, Boulder CO 80309 \\
$^4$ School of Physics, University of Exeter, Stocker Road, Exeter,
 EX4 4QL \\}

\maketitle

\begin{abstract}
We use numerical simulations to model the migration of massive planets
at small radii and compare the results with the known properties
of 'hot Jupiters' (extrasolar planets with 
semi-major axes $a < 0.1$~AU).  For planet masses 
$M_p \sin i > 0.5 M_J$, the evidence for any `pile-up' at 
small radii is weak (statistically insignificant), and although the mass function 
of hot Jupiters {\em is} deficient in high mass planets as compared 
to a reference sample located further out, the small sample 
size precludes definitive conclusions. 
We suggest that these properties are consistent with disc migration 
followed by entry into a magnetospheric cavity close to the star. 
Entry into the cavity 
results in a slowing of migration, accompanied by a growth in 
orbital eccentricity.  For planet masses in excess of $1$ Jupiter
mass we find eccentricity growth timescales of a few $\times 10^5$
years, 
suggesting that these planets may often be rapidly destroyed. Eccentricity 
growth appears to be faster for more massive planets which may explain 
changes in the planetary mass function at small radii and may also
predict a pile-up of lower mass planets, the sample of which is still
incomplete.
\end{abstract}

\begin{keywords}
solar system: formation --- planets and satellites: formation --- 
planetary systems: formation
\end{keywords}

\section{Introduction}
The first extrasolar planet discovered around a Solar-type star, 51~Pegasi~b \citep{mayor95}, 
was the prototype of a class of exoplanets known as `hot Jupiters'. These are massive, gas 
giant planets, which for the purposes of this paper we define to orbit within $0.1$~AU 
of their parent star. Although it has long been clear that the early discovery of 51~Peg 
was due to observational selection effects, and that the typical extrasolar planet orbits 
much further out \citep{marcy05}, hot Jupiters remain of particular interest for two 
reasons. First, the high temperatures of the protoplanetary disc at the radii where 
hot Jupiters now orbit \citep{bell97} mean that it is extremely unlikely that these 
planets formed {\em in situ}. Rather, they must have formed at larger radii and then 
migrated inward to their current location. Although the same is probably also true of 
all extrasolar planets within the snow line \citep{garaud07}, at larger radii the 
requirement for migration is a model-dependent statement whose validity rests on 
the accuracy of giant planet formation models \citep{bodenheimer00}. Second, 
hot Jupiters orbit close enough that interactions with the star cannot be 
ignored. This makes them valuable testbeds for theories of tidal interaction  
\citep{ogilvie04,ivanov07} and planetary structure in the presence of 
irradiation \citep{burrows07}.

The best-developed model for the origin of hot Jupiters and other short-period 
planets attributes their migration to angular momentum loss to gas in the 
protoplanetary disc \citep{goldreich80,lin86}. In this paper we consider 
primarily relatively massive planets, which 
are expected to migrate within gaps in the Type~II regime\footnote{The 
lowest mass short-period planets --- `hot Neptunes' \citep{santos04,mcarthur04} --- 
may instead fall into the gap-less Type~I regime \citep{ward97}. Their origin 
poses different problems that we do not consider here.}. Observations provide 
two apparently contradictory constraints on such models. First,  
the radial distribution of planets is broadly consistent with that predicted by gas disc migration 
models \citep{armitage02,trilling02,ida04,armitage07}. This is true {\em even 
for the hot Jupiters}. In particular there is no substantial pile-up of planets at 
small radii that would be indicative of an efficient mechanism stopping planets 
migrating into the star. Second though, the inner edge of the hot Jupiter 
distribution does not appear to coincide with the Roche limit, as might be 
expected if the current population of hot Jupiters is a remnant of a much
bigger initial population that migrated inward on circular orbits. Rather, 
the inner edge for hot Jupiters occurs at a distance nearly twice that of the Roche limit, 
consistent with circularization from high eccentricity orbits  
in a planet-planet scattering model \citep{faber05,ford06}. To explain this 
within a disc migration model, we require a mechanism that depletes planets 
at radii beyond the Roche limit, but which does not slow migration so much 
as to yield far more hot Jupiters than are observed.

In this paper, we study the dynamical evolution of massive planets as they 
migrate into gas-poor magnetospheric cavities surrounding young stars. A 
large body of observational evidence suggests that the magnetic fields of 
T Tauri stars are typically strong enough to disrupt the inner regions of 
protoplanetary discs \citep{konigl91,bouvier07}, and the existence of this 
inner cavity will affect planet migration if the size of the cavity exceeds 
the stellar Roche limit. It is expected that the stellar magnetosphere will 
truncate the gas disc close to the co-rotation radius, where the Keplerian
angular velocity of the disc is the same as the stellar angular velocity. 
Observationally, pre-main sequence stars have a bimodal distribution of 
rotation periods \citep{herbst07}, with one peak at periods of $\sim 1$ day 
and a second at periods of $5 - 8$ days. Migration is expected to slow 
dramatically once planets pass inside the 2:1 resonance with the inner disc 
edge, so the existence of this long-period population appears consistent 
with hot Jupiters having orbital periods of 3 days or more \citep{lin96}. 
Indeed, given the prevalence of large magnetospheric cavities, the surprise 
is not that there are so many hot Jupiters, but rather {\em why are there 
so few}? In this paper, we argue that the answer lies in the fact that 
entry into the magnetosphere not only slows radial migration, but also 
excites orbital eccentricity. Since the rate of migration and the rate 
of eccentricity growth are directly linked (both are proportional to the 
gas surface density at the inner edge of the disc), planets entering the 
magnetosphere are destroyed as a consequence of eccentricity growth on 
a similar timescale to that on which they would accumulate at small 
radii. An inner edge to the hot Jupiter distribution is created without 
any substantial pile-up of planets there.

The plan of this paper is as follows.  
In Section~2 we describe our model for planet evolution within 
magnetospheric cavities. Results are presented in Section~3, and
Section~4 summarizes our conclusions.

\section{Planet migration within stellar magnetospheres}
Accretion onto low-mass, pre-main sequence stars, known as T
Tauri stars, is thought to occur via accretion along the
stellar magnetic field lines. Since the accreting material is
initially spiralling inwards through a circumstellar disc, the
general idea \citep{ghosh79} is that the disc will be truncated
at the radius at which the ram pressure of the accreting material 
balances the magnetic pressure of the stellar magnetic field. The 
material will then follow the field lines, giving rise to accretion
streams and creating accretion shocks near the stellar magnetic poles. 
For accretion to occur, the truncation radius must be inside co-rotation, 
since outside co-rotation any material tied to the stellar
magnetic field lines will gain angular momentum and will tend to
move away from the star. At the late epochs relevant for studies 
of planet migration the accretion rate is low, so the most 
likely scenario is that the the truncation radius is very  
close to the co-rotation radius. Stellar angular momentum 
considerations \citep{cameron93} also support the idea that 
the disc does not penetrate far inside corotation. Discs around 
T Tauri stars with rotation periods between $5$ and $8$ days
would therefore be truncated at between $0.06$ and $0.08$ AU.   

While the gas disc is still present, planets will migrate inwards
via either Type I \citep{ward97} or Type II \citep{goldreich80} 
migration, depending on the mass of the migrating planet. Once inside
co-rotation, and hence once inside the disc truncation radius, the
planet migration should be quite different.  Type I migration can
no longer operate since there will be no mass at the locations of the
nearby resonances to supply the differential Lindblad torques
\citep{ward97}. Type II will also not operate in the usual manner 
since the standard scenario is that disc viscosity
causes material to flow into the gap, causing the gap and planet
to move inwards on viscous timescales \citep{lin86}. Once the
planet is inside co-rotation, however, the disc material
flows along field aligned accretion streams, accreting onto the
star near the magnetic poles, and does not flow into the gap. 
The low density gas within the magnetosphere does not 
materially affect the planet's orbit \citep{papaloizou07}. 
The planet can, however, interact gravitationally with the
surrounding disc, and it this process that we will investigate
here.

\subsection{Numerical simulations}
We use the Zeus code \citep{stone92} to simulate the circumstellar
gas disc. We work in two-dimensional cylindrical polar coordinates ($r, \phi$) with
a resolution of $n_r = 400$ and $n_\phi = 400$. The computational
domain extends, in code units, from $r_{\rm in} = 1$ to 
$r_{\rm out} = 10$, and we impose outflow boundary conditions 
at both $r_{\rm in}$ and $r_{\rm out}$.  We assume the 
inner boundary is located at the corotation radius and so for
a solar mass star with a rotation period of $8$ days, the inner boundary is
then located, in real units, at $r = 0.08$ AU. The use of an outflow
boundary condition here then approximates the gas accretion onto
the central star.

We assume that the disc is isothermal with a radially dependent
sound speed given by $c_{\rm s}(r) = 0.05 r^{-1/2}$. The disc thickness
$h/r$ is related to the sound speed through $h/r \approx c_{\rm
  s}/V_{\rm K}$, giving a disc thickness of 0.05 at $r = 1$.  We model
angular momentum transport using a kinematic viscosity, $\nu$, that 
operates only on the azimuthal component of the momentum equation
\citep{papaloizou86}. The initial disc surface density, $\Sigma$, 
is taken to have a radial dependence of $\Sigma(r) \propto r^{-1}$. 
Since $\dot{M} \propto \nu \Sigma(r)$ and since we expect $\dot{M}$ 
to be constant, we assume that $\nu(r) \propto r$ and normalise the
viscosity using the standard alpha formalism, $\nu = \alpha c_{\rm s}
h$ \citep{shakura73} with $\alpha = 10^{-3}$ at $r = 1$.

We adopt units in which $G = 1$ and the mass of the star and planet
satisfy $M_* + M_{\rm pl} = 1$. We consider various planet masses and positions,
but the planet is always located within $r = 1$, and the centre
of mass is fixed at $r = 0$.  The dynamics of the star and planet      
are integrated using a second-order scheme that is adequate for
the relatively short durations (a few hundred planetary orbits)
of our simulations. At the epoch of giant planet migration we
would expect the surface density at $\sim 0.1$ AU to be 
$\sim 1000$ g cm$^{-2}$ \citep{armitage96}. This would produce a physical migration
timescale, for hot Jupiters, of $\sim$ 10$^4$ years, corresponding to
around
$10^6$ orbital periods at these small radii.  We are able to evolve
our simulations for only a few hundred orbital periods, and hence this
surface density would not produce measurable changes in the planet's
semi-major axis $a$ or eccentricity $e$ during the course of our
simulations.  To see evolution within
a few hundred orbits we adopt a higher scaling for the surface 
density. This, however, does not affect the rate of eccentricity
growth, measured as $a {\rm d} e/ {\rm d} a$, because the 
accelerations in $a$ and $e$ are linear in disc mass \citep{armitage05,
artymowicz91}.  We therefore use, in code units, $\Sigma = 0.01$ 
at $r = 1$. This corresponds, for a solar mass star with an 8 day rotation
period, to a surface density of $1.4 \times 
10^7$ g cm$^{-2}$, a value $\sim 10^4$ times greater than what one would
realistically expect, but which allows us to measure the rate of change of 
$a$ and $e$ in our simulations.

The simulations consider a star with mass $M_*$ and planet with mass
$M_{\rm pl}$ in orbit with semimajor axis $a$ ($a < 1$) and initial
eccentricity $e = 0.02$.  The disc is initially azimuthally uniform and 
in Keplerian rotation about the centre of mass located at $r = 0$. In
the
presence of a planet, the disc surface density has a non-axisymmetric
distribution that is approximately static in the corotating frame. At
the start of the simulations, there is a transient phase as the disc
adjusts
from the axisymmetric initial conditions.  These transients have no
physical relevance, as real planet formation time scales are long compared
to the orbital period.  To avoid spurious migration in the transient
phase,  we evolve the simulations for $40$ orbits at $r = 1$ 
without the disc mass being included in the calculation of the 
accelerations of the star and planet.  After $40$ orbits, the disc
mass is turned on smoothly over $5$ additional orbits.

\section{Results}
The initial simulations considered planet masses of 
$M_{\rm pl} = 0.001$ and $M_{\rm pl} = 0.01$ and initial semimajor 
axes of $a_{\rm o} = 0.9$, $a_{\rm o} = 0.8$, $a_{\rm o} = 0.7$, 
and $a_{\rm o} = 0.6$. In all the simulations,
the star's mass is set such that $M_* + M_{\rm pl} = 1$. If we assume
the central star has a mass of $1$ solar mass, these two 
planets masses are roughly
$1$ and $10$ Jupiter masses respectively.  All of the simulations
were evolved for at least $250$ orbits at $r = 1$.  As discussed
above, during the first $40$ orbits, the disc mass is not included 
in the star-planet orbit integration and is then turned on smoothly 
over the next 5 orbits.
  
\begin{figure}
{\psfig{figure=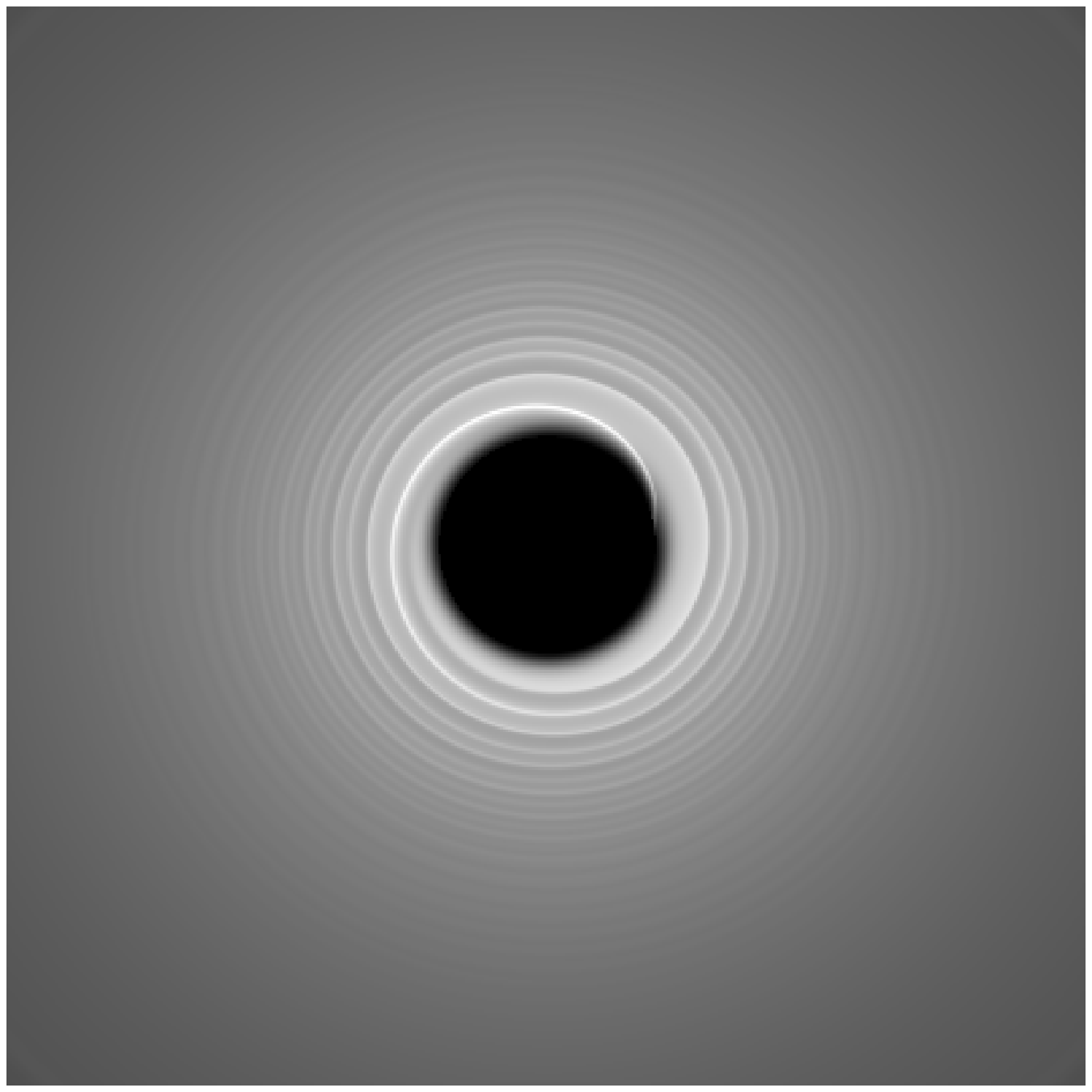, width=41mm}}
{\psfig{figure=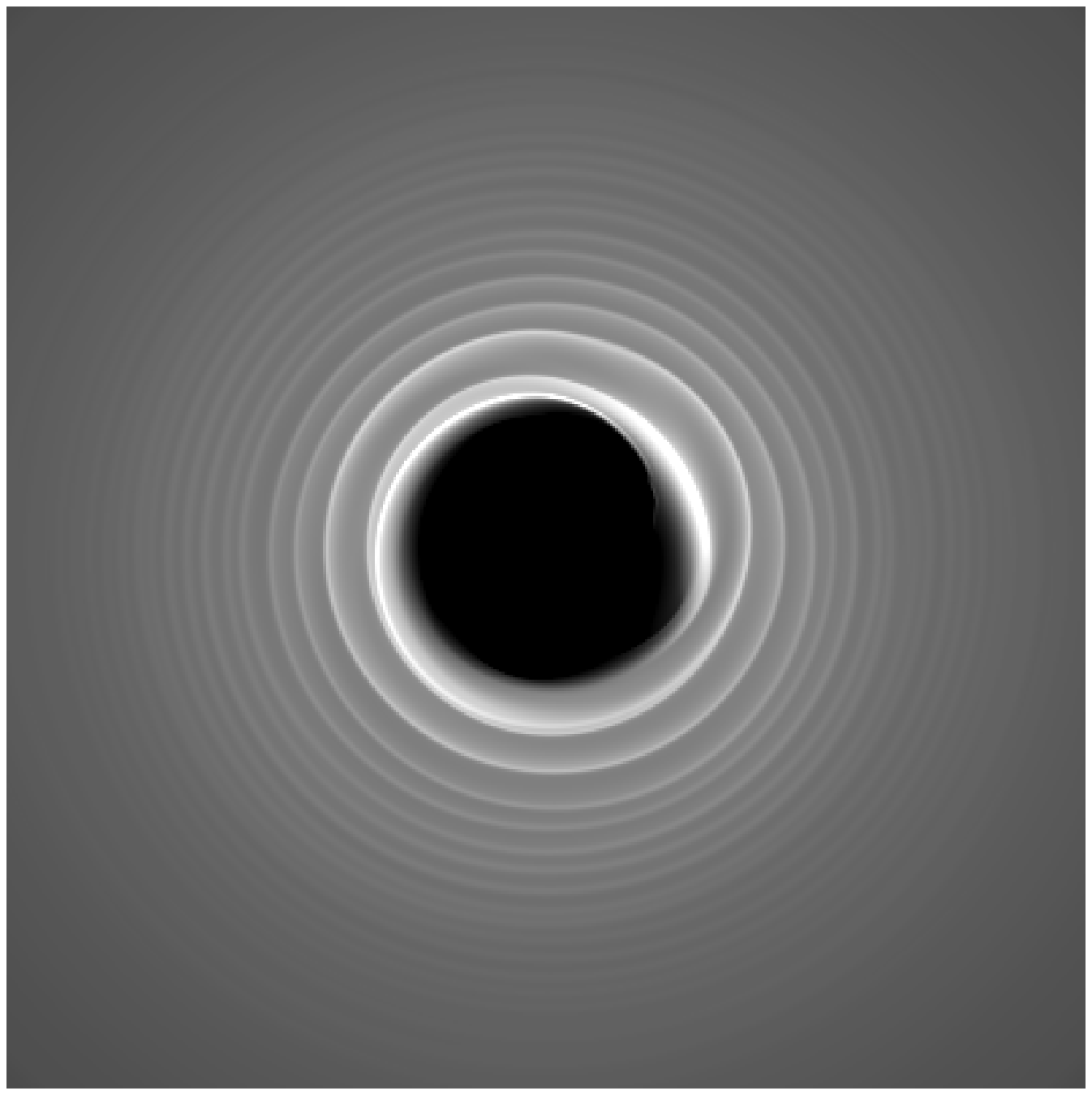,width=41mm}}
\caption{Disc surface density structure 
for $M_{\rm pl} = 0.001$ (left hand panel)
and $M_{\rm pl} = 0.01$ (right hand panel) immediately prior to including 
the disc mass in the calculation of the star and planet accelerations.  
The centre of each image is located at the centre of mass of the star planet
system, and the images extend to $r = 5$.  In both images $a = 0.9$. Although the images
are similar, the spiral density
waves are stronger and extend further into the disc for $M_{\rm pl} = 0.01$ than
for $M_{\rm pl} = 0.001$. }
\label{surfdenstruct}
\end{figure}

Figure \ref{surfdenstruct} shows the surface density structure of
the disc just before including the backreaction of the disc on the
star-planet orbit for $M_{\rm pl} = 0.001$ (left hand panel) and 
$M_{\rm pl} = 0.01$ (right hand panel).  The centre of the images is 
located at the centre of mass of the star planet system and the images
extend to $r = 5$. In both
images the semimajor axis of the star-planet system is $a = 0.9$.  
Although similar, it is clear that the spiral arms are stronger and 
extend further into the disc for $M_{\rm pl} = 0.01$ than 
for $M_{\rm pl} = 0.001$. 
  
\subsection{$M_{\rm pl} = 0.001$}
\begin{figure}
{\psfig{figure=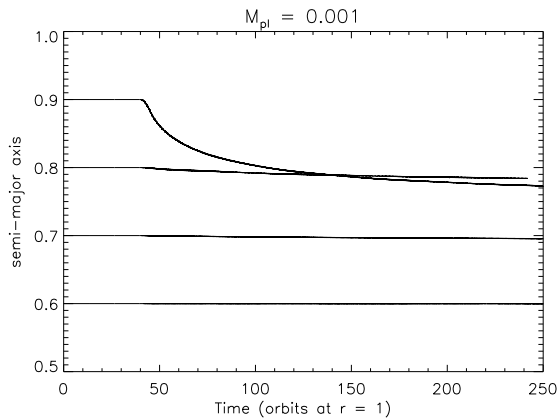, width=80mm}}
\caption{Semi major axis evolution for a planet mass
of $M_{\rm pl} = 0.001$ starting at $a_o = 0.9, 0.8,
0.7, {\rm and} \  0.6$. The decay rate is quite substantial for
$a_o = 0.9$, but decreases significantly with decreasing
initial semi major axis.  There is no noticeable
inward migration for $a_o = 0.6$.}
\label{semimajor_1MJ}
\end{figure}

\begin{figure}
{\psfig{figure=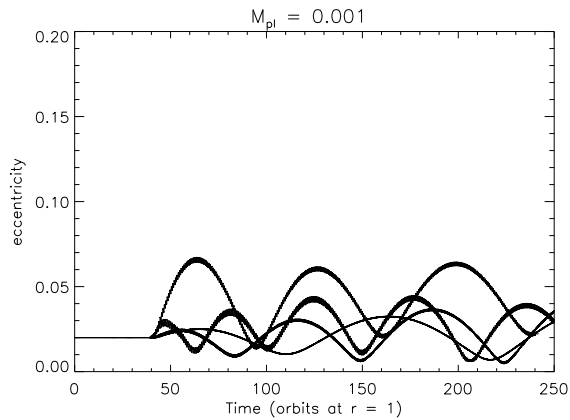, width=80mm}}
\caption{Eccentricity evolution for $M_{\rm pl} = 0.001$ and
for $a_o = 0.9, 0.8, 0.7, {\rm and} \ 0.6$. The eccentricity has
been averaged over the orbital period to remove small scale
flucatuations.  Although there
are some sinusoidal variations in the eccentricites, 
the mean eccentricity does
not appear to change significantly in any of the 4 simulations.}
\label{eccgrowth_1MJ}
\end{figure}

The simulations all start with a small eccentricity ($e = 0.02$) and
consider semimajor axes of $a_{\rm o} = 0.9$, $a_{\rm o} = 0.8$, 
$a_{\rm o} = 0.7$, and $a_{\rm o} = 0.6$. Figure 
\ref{semimajor_1MJ} shows the semimajor axis evolution for 
$M_{\rm pl} = 0.001$.  Apart from the $a_{\rm o} = 0.9$ case, the change in
semimajor axis once the backreaction of the disc is included is very 
small, with absolutely no noticeable change for $a_{\rm o} = 0.6$.  The figure
does show the planet starting at $a = 0.9$ migrating in to a smaller radius than the planet
starting at $a = 0.8$.  This is simply because when the planet starting at $a = 0.9$ reaches
$a = 0.8$ it does not have exactly the same properties as the planet starting at $a = 0.8$, and therefore
does not evolves in exactly the same way.  Its migration does, however, slow considerably once
inside $a = 0.8$, and its inward migration rate approaches a value similar to that of the
planet starting at $a = 0.8$. 
The eccentricity is shown in Figure \ref{eccgrowth_1MJ}.  Although we 
haven't labelled the individual curves, it is clear that apart from
some sinusoidal variations,
the mean eccentricity does not appear to vary significantly over the course of
these simulations.  These results suggest that planets of this mass
will migrate rapidly inwards to $r \sim 0.8$ and then continue migrating more 
slowly, possibly becoming stranded at a radius of between $r = 0.7$
and $r = 0.6$. 

We can also quantify the migration rate in these simulations.
Assuming $r = 1$ corresponds to a real radius of $0.08$ AU,
our simulations are evolved for  $4.6$ years after the backreation of the disc is
included (i.e., from $t = 40$ orbits to $t = 250$ orbits with each 
orbit taking $8$ days). The planet starting at $a_o = 0.9$ therefore
actually starts at $0.072$ AU and migrates inwards to $0.062$ AU in $4.6$ years. Our 
chosen surface density is, however, also about 10000 times greater than would be 
realistically expected.  Since the migration
rate depends linearly on the disc surface density \citep{armitage05,
artymowicz91} a more realistic migration timescale in this case
is $0.015$ AU in $46000$ years, giving a migration rate of $\dot{a} = 3.3 \times 10^{-7}$ AU/yr.  
Repeating this for $a_o = 0.8$ gives
a migration rate of $\dot{a} = 2.8 \times 10^{-8}$ AU/yr. This is clearly
small, but at this rate the planet could still move inwards by $0.01$
AU in $\sim 3 \times 10^5$ years.  For $a_o = 0.7$ and $a_o = 0.6$,
the unscaled migration rates are $\dot{a} = 7 \times 10^{-9}$ AU/yr
and $\dot{a} = 8 \times 
10^{-10}$ AU/yr. Although a planet starting at $a_o = 0.7$ could migrate a
further $0.01$ AU in the disc lifetime (i.e., within a few million
years), this is not possible for $a_o = 0.6$, requiring in excess
of $10^7$ years to do so.  
  
\begin{figure}
{\psfig{figure=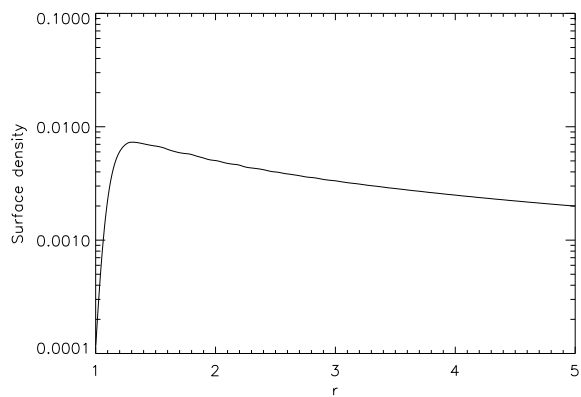, width=80mm}}
\caption{A representative, azimuthally averaged surface density profile, in this case
from a simulation in which
$M_{\rm pl} = 0.001$ and $a_o = 0.9$.  Although the disc
inner edge is located at $r = 1$, the peak of the surface
density occurs at between $r = 1.2$ and $r = 1.3$.  A planet
with mass $M_{\rm pl} = 0.001$ appears to stall just inside the
radius that is in 2:1 resonance with the peak of the surface
density.}
\label{sdens1MJ}
\end{figure}

In the above case, the angular momentum exchange between the planet 
and disc is likely
to occur primarily through the 2:1 resonance. Figure \ref{sdens1MJ} 
shows a representative, azimuthally averaged disc surface density profile just prior to 
the start of active integration. Although the inner edge of
the disc is at $r = 1$, the peak of the surface density occurs at 
between $r = 1.2$ and $r = 1.3$. The 2:1 resonance with the peak of 
the surface density therefore occurs at between $r = 0.76$ and 
$r = 0.82$. The interpretation of the above result is that as the planet
migrates inwards, the location of the 2:1 resonance moves inside the peak of the
surface density, the angular momentum exchange becomes inefficient,
and the migration stalls inside, but reasonably near, the
radius that is in 2:1 resonance with the peak of surface density. This
gives an orbital period for the planet that is close to half the
rotation period of the central star.

\subsection{$M_{\rm pl} = 0.01$}
\begin{figure}
{\psfig{figure=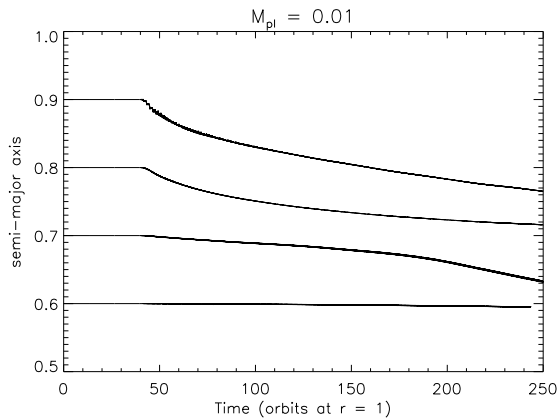, width=80mm}}
\caption{Semi major axis evolution for a planet mass
of $M_{\rm pl} = 0.01$ starting at $a_o = 0.9, 0.8,
0.7, {\rm and} \ 0.6$. The decay rate is quite substantial for
all initial semi major axes except $a_o = 0.6$, which still
has non-negligible inward migration. What is most interesting
is how, for $a_o = 0.7$, the inward migration appears to accelerate after
about 180 orbits.}
\label{semimajor_10MJ}
\end{figure}

\begin{figure}
{\psfig{figure=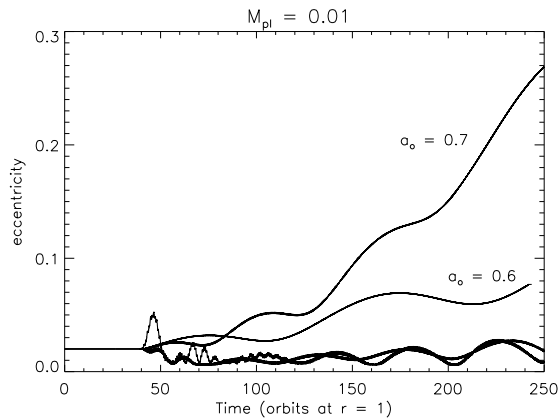, width=80mm}}
\caption{Eccentricity evolution for $M_{\rm pl} = 0.01$ and
for $a_o = 0.9, 0.8, 0.7, {\rm and} \ 0.6$. The eccentricity has
been averaged over the orbital period to remove small scale
flucatutions.  What is evident from the figure is that a planet
starting at $a_o = 0.7$ or $a_o = 0.6$ will undergo substantial 
eccentricity growth, in particular for $a_o = 0.7$ in which the 
eccentricity grows from $e = 0.02$ to $e = 0.27$ during the 
simulation.}
\label{eccgrowth_10MJ}
\end{figure}

We repeat the above simulations, keeping everything the same
except the planet mass which is now $M_{\rm pl} = 0.01$, 10 times 
more massive than that considered above and which corresponds, for our
chosen scaling, to $\sim 10$ Jupiter masses.  Figure 
\ref{semimajor_10MJ} shows the semimajor axis evolution for 
$M_{\rm pl} = 0.01$.  This is considerably different to
the $M_{\rm pl} = 0.001$ case.  All four cases show
inward migration, with the rate being quite substantial for 
$a_{\rm o} = 0.9$, $a_{\rm o} = 0.8$, and $a_{\rm o} = 0.7$.  What is most interesting is that
in the $a_{\rm o} = 0.7$ case, the migration rate appears to accelerate
after $\sim 180$ orbits.  When we consider the eccentricity evolution,
shown in Figure \ref{eccgrowth_10MJ}, we see that in this case
the eccentricity growth is substantial, increasing from $e = 0.02$ 
to $e = 0.27$ during the course of the simulation.  There is still
noticeable eccentricity growth for $a_o = 0.6$, but not much growth
for $a_{\rm o} = 0.9$ and $a_{\rm o} = 0.8$ which are unlabelled in
Figure \ref{eccgrowth_10MJ}.  The angular momentum exchange between
the planet and disc occurs generally through corotation resonances - which in the
case of an eccentric planet may be non-coorbital - and Lindblad 
resonances.  It is fairly well known that Lindblad
resonances produce eccentricity growth, while corotation resonances 
produce eccentricity damping, and that if corotation
resonances are present, they will tend to dominate \citep{goldreich80}. 
A way in which eccentricity growth can
occur, however, is if the gap around a planet is large enough that the
outer
resonance still operates, but the corotation resonances do not 
\citep{dangelo06,artymowicz91}.  Our simulations, therefore, suggest
that when the planet is at large radii ($a = 0.9$, and $a = 0.8$) the 
outer Lindlad resonances do not dominate sufficiently to produce significant
eccentricity growth.  At smaller radii ($a = 0.7$), however, the gap is
effectively larger, the inner resonances no longer operate
efficiently,
and the outer resonances, primarily the 1:3 resonance \citep{dangelo06,artymowicz91}, act to produce
eccentricity growth.  As the planet moves to even smaller radii ($a = 0.6$), 
the outer resonances stop operating efficiently, and the
eccentricity growth rate
starts to decrease.

\begin{figure}
{\psfig{figure=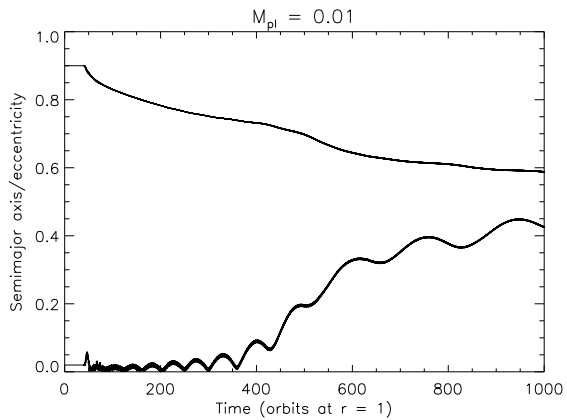, width=80mm}}
\caption{Semimajor axis (upper curve) and eccentricity (lower curve)
evolution for a $M_{\rm pl} = 0.01$ planet starting at $a = 0.9$, and
evolved
for $1000$ orbits. It is clear that when this planet reaches $a \sim 0.7$, 
it undergoes rapid eccentricity growth, the rate of
which
decreases as it moves further inwards, but at the end of the
simulation
is still not negligible.}
\label{longrun}
\end{figure}

These simulations suggest that the more massive planet will
migrate inwards rapidly until reaching $a \approx 0.7$ at which radius
it will start to undergo rapid eccentricity growth. Unlike the 
lower mass case, there is no evidence here that the planet migration
is likely to stall and we see inward migration and eccentricity
growth even for the smallest initial semimajor axis considered. 
Figure \ref{longrun} shows the long term evolution in semimajor axis and
eccentricity for a $M_{\rm pl} = 0.01$ planet starting at $a = 0.9$.
The figure shows that the initially the eccentricity growth is
small
but that the rate of eccentricity growth increases rapidly when $a
\sim 0.7$.  As the planet continues migrating inwards the eccentricity
growth rate does decrease, but at the end of the simulation the planet
is still migrating inwards and its eccentricity is still increasing.
If we again assume our inner disc radius corresponds to a real
radius of $0.08$ AU, at the end of this simulation the planet will have a 
periastron distance of only $5.5$ solar radii. The simulation is
evolved
for $1000$ orbits at $r = 1$ which, if we assume a radial scaling of 
$0.08$ AU, corresponds to a time of about $4000$ days.  As discussed
earlier, our chosen surface density is about $10000$ times
greater than a realistic surface density.  A more realistic timescale
for this evolution is thefore $\sim 10^5$ years, which is well within
the
disc lifetime. Since young stellar objects generally have radii of a few solar
radii, and since this planet is still migrating inwards at the end
of this long simulation, it seems quite likely that it will ultimately
collide with the central star.

\subsection{$a_{\rm o} = 0.7$}
To further investigate how eccentricity growth varies with planet
mass, we have considered additional planet masses of 
$M_{\rm pl} = 0.005$, and $M_{\rm pl} = 0.0005$, corresponding for
our chosen scaling, to
roughly $5$, and $0.5$ Jupiter mass planets, both with 
initial semimajor axes of $a_{\rm o} = 0.7$. This initial
semimajor axis is chosen because it 
showed the greatest eccentricity growth in the $M_{\rm pl}
= 0.01$ simulations.   

\begin{figure}
{\psfig{figure=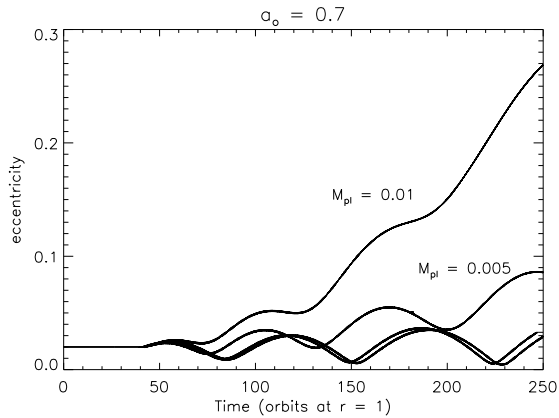, width=80mm}}
\caption{Eccentricity growth for planet masses of $M_{\rm pl} =
0.01$, $M_{\rm pl} = 0.005$, $M_{\rm pl} = 0.001$, and $M_{\rm pl} 
= 0.0005$ all starting at $a_o = 0.7$.  For our chosen scaling these correspond to planet
mass of $10$, $5$, $1$, and $0.5$ Jupiter masses. What is clear is 
that the eccentricity growth rate decreases with decreasing planet
mass.  For realistic surface densities and assuming an inner disc
radius at $0.08$ AU, the two most massive planets can undergo
substantial
eccentricity growth in a time of $\sim 10^5$ years.}
\label{eccen_10-05MJ}
\end{figure}

Figure \ref{eccen_10-05MJ} shows the eccentricity growth for all
the simulations starting with $a_{\rm o} = 0.7$.  The 
$M_{\rm pl} = 0.01$ and $M_{\rm pl} = 0.005$ cases are labelled,
while the eccentricity evolution for the other 2 planet masses 
($M_{\rm pl} = 0.001$ and $M_{\rm pl} = 0.0005$)
are very similar and can't be easily distinguished.
What is clear is that the eccentricity growth rate decreases
with decreasing mass. Fitting straight lines to
the eccentricity curves between $t = 40$ orbits and $t = 250$ 
orbits we find, for our chosen scaling, eccentricity
growth rates, in order of decreasing planet mass, of $\dot{e} = 4.8 \times 
10^{-6} {\rm yr}^{-1}$, $\dot{e} = 1.3 \times 10^{-6} {\rm yr}^{-1}$,
$\dot{e} = 1.0 \times 10^{-7} {\rm yr}^{-1}$, and $\dot{e} = 3.1 \times 10^{-8} 
{\rm yr}^{-1}$. The latter two growth rates should be treated
somewhat cautiously since the error in these values will be large
compared to the values themselves.  These growth rates do, however, 
suggest that planets with
masses greater than $1$ Jupiter mass may undergo substantial
eccentricity growth in a time of $\sim 10^5$ years, while lower mass planets 
may require $10^6$ years or longer. This is at least consistent
with the lack of a pile-up of hot Jupiters at small 
radii, but may predict a pile-up of lower mass
planets, the sample of which today is still incomplete.

It should however be noted
that eccentricty growth rates appear to depend on the disc
viscosity \citep{dangelo06,kley06}. Eccentricity growth can be
sustained in discs with $\alpha$ less than a few times $10^{-3}$
and increases with decreasing $\alpha$,
while $\alpha$ values in excess of $10^{-2}$ may ultimately damp
eccentricity growth \citep{dangelo06}. \citet{moorhead07} similarly
argue that eccentricity growth depends strongly on the properties of
the gap, which is determined both by the disc viscosity and
the planet mass \citep{syer95}. Our simulations involve 
analogous, but not identical considerations.  In our case viscosity
is unimportant within the cavity (by assumption in the simulations,
and also physically), and the gap edge is set by the balance between
viscosity and magnetospheric, rather than gravitational, torques.

\section{Discussion and Conclusion}
In this paper we have considered star-planet
systems with initially small eccentricities and with orbital radii
$a < 1$, surrounded by a circumstellar disc extending from
$r = 1$ to $r = 10$.  The inner edge of the disc is assumed to be
truncated by the stellar magnetic field close to corotation and we
assume that the planet has migrated inwards through the disc 
and is now inside the magnetospheric cavity. We have generally assumed
that the inner edge of our disc corresponds to a real radius of $0.08$
AU, as expected for a stellar rotation period of $8$ days. 

Our initial simulations consider planet masses of $M_{\rm pl} 
= 0.001$ and $M_{\rm pl} = 0.01$ and initial semimajor axes
of $a_{\rm o} = 0.9$, $a_{\rm o} = 0.8$, $a_{\rm o} = 0.7$, 
and $a_{\rm o} = 0.6$.  The $M_{\rm pl} = 
0.001$ simulation, which is taken to represent a $\sim 1$ Jupiter
mass planet, initially migrates inwards rapidly, but appears to 
stall at a radius of between $r = 0.6$ and $r = 0.7$.  This is 
close to the radius that is in 2:1 resonance with the peak
of the disc surface density which in our simulations is just outside
the inner edge of the disc.  Assuming a stellar
rotation period of $8$ days, this suggests that these planets
would stall with orbital periods of $\sim 4$ days.  More generally,
the planet would stall with an orbital period a little less than
half the stellar rotation period.

The $M_{\rm pl} = 0.01$ simulation was significantly different. The 
inward migration was much more rapid and the simulation starting
with $a_{\rm o} = 0.7$ showed significant eccentricity growth. 
To produce observable changes within the few hundred orbits
of our simulations, we used surface densities significantly higher than
would be realistically expected.  After scaling the surface density
to a more reasonable value, this planet would survive $\sim 10^5$ years
before colliding with the central star. To further
investigate this process we repeated this simulation ($a_{\rm o} = 0.7$) but
varied the planet mass from $M_{\rm pl} = 0.01$ down to $M_{\rm pl} =
0.0005$. As expected, the eccentricity growth rate increased with 
increasing planet mass, with the 
lowest mass planet showing
very little or no eccentricity growth, while the eccentricity of the
most massive planet increased substantially.  The
growth rates calculated assuming stellar rotation periods of $8$ days,
and reasonable disc surface densities, suggest that planets with masses
in excess of $1$ Jupiter mass could undergo substantial eccentricity
growth in $\sim 10^5$ years, while the lower mass planets need $10^6$
years or longer. The exact mass range for which we might expect 
substantial eccentricity growth is, however, difficult to accurately
quantify as it depends not only on the disc surface density, but also
on the disc viscosity \citep{dangelo06}. 

These general results of our simulations appear at least qualitatively consistent with
current observational evidence. That eccentricity growth rates
increase
with increasing mass is consistent with the possible deficit of massive
planets at small radii \citep{zucker02} and is also at least 
qualitatively consistent with the lack of any significant pile-up
of giant planets at small radii. These simulations also suggest 
that some planets could be left stranded 
on eccentric orbits if the disc dissipates prior to these
planets colliding with the central star. The decrease in migration
rate with decreasing mass would imply that there should be some
mass range for which this becomes quite likely, i.e., when the 
eccentricity growth timescale approaches the disc lifetime.
\citet{ford06} suggest that if hot Jupiters migrate 
inwards on orbits with reasonably small eccentricities and then
become tidally locked to the central star, the inner edge of the
hot Jupiter population should be close to the Roche limit, the
distance at which a planet fills its Roche lobe.  They find,
however, that the inner edge of the hot Jupiter population
appears to be located at twice the Roche limit, rather than at
the Roche limit itself, as would occur if these planets
are tidally circularised from highly eccentric orbits \citep{faber05}.

What is also interesting is that this does not appear to be true
for the lowest mass close-in planets observed to date, the ``hot
Neptunes'', a few of which appear to exist close
to their Roche limit.  This is again qualitatively consistent
with our result that the lowest mass planets undergo very little
eccentricity growth. The inner edge of this population should
then occur at the Roche limit, rather than at twice the Roche 
limit as would occur if they were tidally circularized from
an eccentric orbit. This result would also suggest that 
we might expect a pile-up of low-mass close-in planets, as their
inward migration appears to stall with a period a little less than
half the stellar rotation period.  

Since the eccentricity growth rate depends on both the disc
mass and viscosity, it is difficult to quantify the exact mass
at which we would expect eccentricity growth to become
negligible.  However, our simulation results do 
suggest three primary outcomes:
\begin{itemize}
\item{Low-mass planets 
migrate inwards on roughly circular orbits and will stall with an
orbital period close to half the rotation period of the central
star. The inner edge of this population should be located near the
Roche limit.  Since these planets do not migrate into the central
star, we would predict a pile-up of low-mass planets at small radii.
}
\item{High-mass planets undergo significant eccentricity growth
and should collide with the central star - for reasonable
disc viscosities and masses - in a time of $\sim 10^5$
years.}  
\item{Intermediate-mass planets, and some high-mass planets,  
could be stranded on eccentric
orbits if the disc dissipates sufficiently quickly. These would
then be tidally circularized and the inner edge of this population
would be located at twice the Roche limit \citep{ford06}.}  
\end{itemize}

\section*{acknowledgements}
This work was supported by NASA's ATP program under grants 
NNG04GL01G and NNX07AH08G, and by the NSF under grant 
AST~0407040.

\end{document}